# Cyclic and tensile deformations of Gold-Silver core shell systems using newly parameterized MEAM force field

Sk Md Ahnaf Akif Alvi[1], Abrar Faiyad[1], Md Adnan Mahathir Munshi[1], Mohammad Motalab[1,*], Md Mahbubul Islam[2], Sourav Saha[1,3,*]

[1]Department of Mechanical Engineering, Bangladesh University of Engineering and Technology, Dhaka-1000, Bangladesh.

[2]Department of Mechanical Engineering, Wayne State University, Detroit, MI 48202, USA.

[3]Theoretical and Applied Mechanics, Northwestern University, Evanston, Illinois, USA.

## Abstract

Gold-Silver (Au-Ag) core-shell nanostructures are gaining importance in stretchable and biocompatible electronics where high tensile and fatigue resistance is of paramount importance. These materials undergo high dislocations, twin planes and stacking fault formations, propagation and annihilation during tensile and fatigue loading. This work, for the first time, quantitatively investigates the role of dislocations and defect interaction governing the mechanical behavior of Au-Ag and Ag-Au Core-shell nanostructures under tensile and fatigue loading using molecular dynamics (MD) simulation. For accurate representation of the underlying physics, a novel modified embedded atomic model (MEAM) interatomic potential is parameterized through density functional theory (DFT) calculations. A comparative analysis between the CSS and their pristine counterparts is also conducted. Throughout this work, pseudo-potential and all electron full potential DFT schemes are used for parameterizing MEAM by calculating cohesive energy, lattice parameter and bulk modulus of pure Au, Ag and their alloy. Using the new force-field for MD simulations, the tensile behavior of pristine and core-shell nanowires is explored for temperatures between 300K to 600K. The fatigue properties of two pristine and two core-shell nanowires in a strain range of -15% to 15% for 10 cycles is also conducted. Our results suggest that Ag-Au Core-shell nanowire show the best reversibility under fatigue loading among the structures examined. Moreover, Ag-Au exhibit highest dislocation formation and complete annihilation of defects consistently. While, Au-Ag present improved fatigue properties than its pristine counterparts but have some residual defects leading to lower reversibility when compared to Ag-Au. For tensile loading, all four structures exhibited deterioration in strength with increasing temperature. Thermal softening is seen to be more prominent in Au-Ag core-shell nanowires compared to Ag-Au. Our work lays out a foundation for exploration of mechanical properties of Au-Ag systems using the MEAM potential which will help design better components for stretchable electronics and creates a pathway for further exploration of similar binary alloy systems.

*Corresponding authors: E-mail address: sourav023@u.northwestern.edu, abdulmotalab@me.buet.edu.bd

# 1. Introduction

Gold-Silver Core-Shell structures (CSS) are becoming increasingly important due to their potential application in stretchable electronics ushering a new era in consumer and bio-electronics, unraveling new branches of material research. From foldable displays and curved solar panels to new classes of soft bio-electromechanical system, stretchable electronics have enabled us to develop flexible electronic/optoelectronic devices, sensors and soft biocompatible instruments(Honda et al., 2014; Hong et al., 2019; Lee et al., 2021; Liu et al., 2017; Lu and Kim, 2014; Takei, 2018). Hence, research on materials with good fatigue resistance, high electric conductance and biocompatibility have become a necessity(Choi et al., 2018; Wang et al., 2022).

Core-shell nanowires (CSNW) and nano-clusters fulfill these stringent requirements while extending the design space of stretchable materials by adding new parameters for tuning targeted propertie(Ah et al., 2001; Hu et al., 2011a; Sunwoo et al., 2020). It has established its place across many diverse fields, like- plasmonics, bioelectronics, and soft-electronics(Brown et al., 2011; Byers et al., 2015; Gontero et al., 2017; Kheradmand et al., 2020; Li et al., 2021; Loiseau et al., 2019; Wang et al., 2011, 2021). Au-Ag nanowires (NWs) and CSNWs exhibit high-stretchability, good fatigue resistance, electrochemical stability, good electric conductance, biocompatibility and optical transparency in polymeric or elastomeric substrate(Choi et al., 2018; Sunwoo et al., 2020). This makes them a prime candidate for stretchable electronics application.

Pure Au, Ag and their alloys are being widely used for different biological and catalytic applications. Ag-Au CSNW composites are being used in bioelectronic devices for biosensing and stimulating human skin and swine heart(Choi et al., 2018). Sung-Hyuk Sunwoo et al. proposed the use of Ag-Au core-shell NW for enhancing pumping capacity of weak cardiac tissue(Sunwoo et al., 2020). The advancements in manufacturing techniques have made manufacturing of Au-Ag CSNWs more effecient(Ah et al., 2001; Choi et al., 2018; Liu and Guyot-Sionnest, 2004; Okuno et al., 2010; Xiang et al., 2008; Yang and Chang, 2006). Au-Ag alloys are used for bio-imaging(Le Guével et al., 2012; Retnakumari et al., 2009; Zerda et al., 2015), tissue engineering and regenerative medicines(Vial et al., 2017), diagnosis(Halo et al., 2014), highly conductive electrical applications(Goodman, 2002; Tickner et al., 2016), pacemaker implantation(Syburra et al., 2010) and Nano-Electro Mechanical Systems (NEMS)(Akbarian and Dehghani, 2020). Catalytic applications such as- heavy ion detection through fluorescence, reaction enhancement of Au-Ag alloys and composites are also noteworthy(Chen et al., 2013; Guével et al., 2011; Haldar et al., 2014; Lates et al., 2014; Liu et al., 2020; Lu et al., 2020; Salehi-Khojin et al., 2013; Sun et al., 2014; Wang et al., 2012; Yang et al., 2016; Young et al., 2016; Zhai et al., 2017; Zhang et al., 2014). In these applications fatigue loading is one of the major causes of material failure during their service life (Akbarian and Dehghani, 2020). The abundance of applications of Au-Ag composites and CSNWs in various fields demands a deeper understanding about the mechanical properties of the material under different loading conditions.

Advancements in computational material science have expedited the modeling of materials with unique properties. Time and cost of synthesis and characterization for such nanostructures are high. Advanced computational techniques, like molecular dynamics (MD), can lower these costs. Studies on the mechanical properties of Cu-Ag, Al-Cu, Si/a-Si, Ni-Co, Cu-Ag CSNWs,(Jing and Meng, 2010; Li et al., 2017; Sarkar, 2018; Sarkar and Das, 2018; Shiave et al., 2019), and thermal properties of Si-Ge CSNWs (Hu et al., 2011a, 2011b) are reported using MD simulations. MD simulations has allowed the investigation of dislocation density, effects of radiation and other properties which are not physically or economically viable for experimental exploration(Islam et al., 2020; Paul et al., 2021). Even the mechanical properties of Au-Ag pristine structures were computed elaborately using MD(Baletto et al., 2000; Chen et al., 2014; Shim et al., 2002; Su et al., 2020; Tian et al., 2008; Yu et al., 2010). For a more detailed understanding of its mechanical properties, computational tools like MD can be employed.

To enable large-scale MD simulations of Au-Ag nanostructures, a reliable interatomic potential (IP) is of paramount importance. A semi-empirical correlation such as IP has to be reliable, flexible, and efficient(Aitken et al., 2021). A reliable IP can accurately reproduce relevant fundamental physical traits, such as- mechanical, structural, and thermal properties of the elements and alloys. The IPs available for Au-Ag alloy systems are not specialized for simulating mechanical response of large scale metallic systems(Titantah and Karttunen, 2013) and do not consider the angular contribution of the atomic bonds(Liu et al., 1991). Modified embedded atomic model (MEAM) is widely accepted as an accurate IP which can model the delocalization of electrons considering the angular contribution of the orbitals and atoms, and can also describe more than one phase, like FCC, HCP, BCC and dimond-structured materials using the same mathematical framework for metals and multi-component alloy systems(Baskes, 1992; Jelinek et al., 2007; Kim et al., 2006; Lee et al., 2003; Lee and Baskes, 2000). This facilitates the description of phase transformations occurring in metals and alloys under deformation. MEAM potential requires a specific set of phenomenological parameters which can be determined through high fidelity methods like density functional theory (DFT) and experimental observations(Kim et al., 2006; Lee, 2006; Yang et al., 2010). DFT has been extensively used for calculation of mechanical, electrical, optical and chemical properties in different material systems for small number of atoms, being computationally expensive(Borges et al., 2010; Chepkasov et al., 2018; Durante et al., 2011; Hernández et al., 2019; Mebi, 2011; Rani et al., 2014; Senanayake et al., 2019; Zhao et al., 2016). But using DFT, a new MEAM potential can be parameterized which can reliably reproduce the mechanical response of large scale Au-Ag systems in MD with less computational resources.

Keeping these scopes in mind, we perform DFT calculations to parametrerize a MEAM force-field for accurate description of Au-Ag nano-structures. To reinforce the applicability of the force-field, we investigate the mechanical properties of Au-Ag and Ag-Au CSNWs under uniaxial tension and cyclic loading conditions and compare them to their pristine NW counterparts using MD. We observe the effects of different temperature between 300K-600K on the NWs and CSNWs under tensile loading. With the help of dislocation dynamics analysis, we also explain the fatigue response of the structures. This work aims to create a pathway for further research on Au and Ag alloys and structures which have immense significance and applications.

# 2. Methodology

## *2.1 Density Functional Theory (DFT) calculations*

We performed DFT calculations using two pseudo-potentials (Ultra-Soft Pseudo-potential (USPP) and Projected Augmented Wave (PAW)) and all-electron full-potential (AEFP) schemes for parameterization of the MEAM force-field(Andersen, 1975; Blöchl, 1994; Kresse and Joubert, 1999; Sjöstedt et al., 2000). The pseudo-potential DFT simulations are executed with Quantum Espresso package(Giannozzi et al., 2009). The Brillouin zone is sampled with k-points using Monkhorst–Pack grid scheme with 10x10x10 mesh(Monkhorst and Pack, 1976). For the AEFP formulation, Exciting (Nitrogen) package is used with a smaller k-point grid of 8x8x8, due to the higher computational cost of AEFP(Gulans et al., 2014). The PBE and PBEsol Generalized Gradient Approximation (GGA) type exchange-correlation functional is used for the Pseudo-potential schemes and the AEFP scheme, respectively(Perdew et al., 1997). The Exciting Nitrogen package implements muffin-tin radius approximation with linearized augmented plane wave methods(Andersen, 1975; Gulans et al., 2014; Sjöstedt et al., 2000). In both Pseudo-potential and AEFP schemes, to calculate cohesive energy is calculated from the equation–

$$E_{cohesive_{AB}} = \left[ E_{AB_{bulk}} - \left( \frac{N}{2} E_{A_{atom}} + \frac{N}{2} E_{B_{atom}} \right) \right] / N \qquad (1)$$

$$E_{cohesive_{A(B)}} = \left[ E_{A(B)_{bulk}} - \left( N E_{A(B)_{atom}} \right) \right] / N \qquad (2)$$

where $E_{AB_{bulk}}$ and $E_{A(B)_{bulk}}$ is the energy of the bulk alloy and pristine structures where A and B respectively denotes Au and Ag atom species. $E_{A_{atom}}$ and $E_{B_{atom}}$ denotes energy of one atom. $E_{A(B)_{atom}}$ values are calculated through Self-Consistent Field (SCF) iteration of a single atom in a large simulation cell to mitigate periodic boundary condition effects.

The SCF scheme calculates energy iteratively by solving simplified many-body Schrodinger's equation, namely the Kohn-sham equation(Hohenberg and Kohn, 1964; Kohn and Sham, 1965). The SCF calculations are done with varying lattice parameters. The energy of the configurations are then plotted against volume and fitted to an Equation of State (EOS).

We calculate the lattice parameter and bulk modulus from a plot of calculated energy against the lattice parameters fitting against Murnaghan EOS(Murnaghan, 1944). The lowest energy configuration provides the equilibrium lattice constant.

### 2.2 MEAM force field parameterization

The lattice parameter, bulk modulus, cohesive energy, and thermal expansion coefficient (TEC) of Au-Ag alloys are calculated using Large-scale Atomic/Molecular Massively Parallel Simulator (LAMMPS)(Plimpton, 1995) with the proposed MEAM potential. For MEAM parameterization- lattice parameter, cohesive energy, and bulk modulus are determined using DFT, and the TEC of Au-Ag alloys are matched with the experimental values. The lattice parameter is calculated from MD simulations using a sample of the alloy of cubic shape with sides of two unit cells (Figure 1 (b)). The structure is equilibrated using conjugate gradient minimization technique. The lattice parameter is then obtained from the final length of one side of the cubic body. The cohesive energy is determined by dividing the total energy of the equilibrated sample by the total number of atoms. The TEC is calculated by using a cubic sample measuring 8-unit cells per side. This sample is then equilibrated and subjected to a temperature increase from 100K-600K under NPT ensemble with Noose-hover thermostat(Evans and Holian, 1985). The length of the equilibrated sample is recorded and plotted against temperature. The TEC is calculated by dividing the slope of the plot by the initial length. For the calibration of the $C_{min}$ and $C_{max}$ of the alloy combinations, the MD simulations of TEC are checked against corresponding experimental values. The Young's modulus of the alloy specimen is calculated from the stress-strain curves.

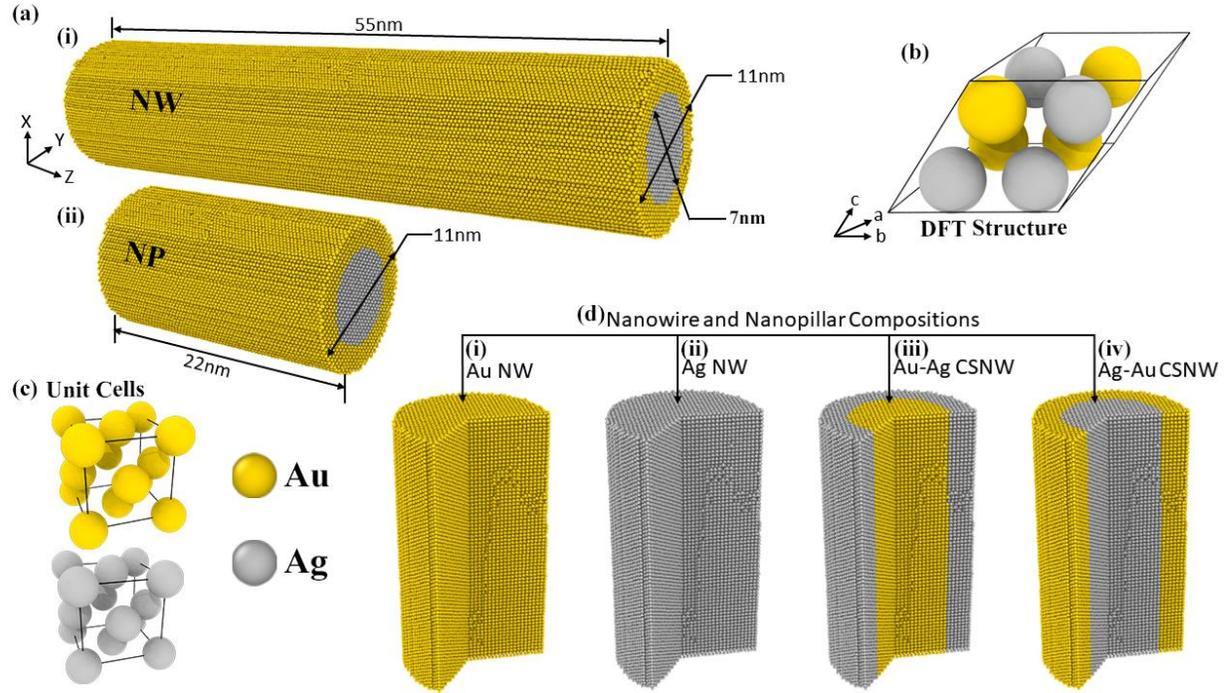

Figure 1: The distribution of Au and Ag in the different structures studied. Here (a) (i) represents the core-shell NW and (ii) represents the core-shell nanopillar structure which is used for simulation. (b) Is the primitive FCC unit cell of with 50% Au and 50% Ag used in pseudo-potential and all-electron full-potential DFT schemes. (c) Shows the conventional FCC unit cells of pristine Au and Ag used for formation of (d) the different core-shell composition- (i) Au, (ii) Ag NWs and, (iii) Au-Ag, (iv) Ag-Au CSNWs used to study the material behavior under tensile and cyclic loads.

### 2.3 Computational Method for MD Simulation

MD simulations are carried out using the LAMMPS and visualized using OVITO(Stukowski, 2009). Two pristine NWs- Au, Ag, and two CSNWs- Au-Ag, Ag-Au are constructed for the simulations to determine the tensile and fatigue properties (Figure 1). We used <-1 1 0>, <0 0 1>, <1 1 0> oriented NWs for this study. For the tensile test, NWs of 11nm diameter and 1:5 diameter-to-length ratio are constructed. Velocity verlet algorithm is used as an integrator with a time step of 1fs. Energy equilibration is done using NVE ensemble with a Langevin thermostat for 50 picoseconds(Schneider and Stoll, 1978). Then the structures are relaxed in NPT ensemble with Noose-hover barostat and NVT ensemble for 50 picoseconds each(Evans and Holian, 1985; Parrinello and Rahman, 1981). After equilibration, with a strain rate of $10^9$ s$^{-1}$ in NVT ensemble tensile behavior is observed up to 20% strain(Parrinello and Rahman, 1981). For fatigue tests, structures of similar diameter and 1:2 diameter-to-length ratio are used. The aspect ratio is decreased to prevent buckling during compression. An equilibration method similar to the tensile test is applied here. The structure is stretched and compressed to 15% of its original length to incorporate plastic deformation in the simulations. The time of one tension-compression cycle was 200 picoseconds. Fatigue properties of the structures are observed for 10 cycles. During tension and fatigue, virial stresses are calculated using the equation-

$$\sigma_{virial}(r) = \frac{1}{\Omega}\sum_i[(-m_i\dot{u}_i \otimes \dot{u}_i + \frac{1}{2}\sum_{j \neq i} r_{ij} \otimes f_{ij})] \tag{4}$$

where the sum is taken for all the atoms in the volume, $m_i$ denotes the mass of atom $i$, $\dot{u}_i$ is the time derivative of the displacement, $r_{ij}$ denotes the position vector, and $f_{ij}$ denotes the interatomic force applied on atom $i$ by atom $j$.

# 3. Results and Discussions

## 3.1. MEAM force field parameterization

Accurately describing a pure element using the 2NN MEAM potential formalism requires 14 independent parameters(Kim et al., 2006). Four of these parameters for pure elements are cohesive energy E$_c$, equilibrium nearest-neighbor distance r$_e$, bulk modulus (B) of the reference structure, and the adjustable parameter d and is related to the universal equation of state. Other parameters [the decay lengths $\beta_i^{(h)}$ (h = 0–3) and the weighting factors $t_i^{(h)}$ (h = 1–3)] are required to compute the electron density. The parameter A belongs to the embedding function, and the parameters $C_{min}$ and $C_{max}$ describe the many-body screening. These 14 parameters for the case of pure Au and Ag was parameterized by Lee et al.(Lee et al., 2003). The parameters for pure Ag and Au as taken from Lee et al. are given below-

Table 1: Parameters for the MEAM potential of Ag and Au. The units of the sublimation energy $E_c$, the equilibrium nearest-neighbor distance $r_e$, and the bulk modulus B are eV, A, and $10^{12}$ dyn/cm$^2$, respectively.

|    | $E_c$ | $r_e$ | $\alpha$ | A | $C_{min}$ | $C_{max}$ | $\beta^{(0)}$ | $\beta^{(1)}$ | $\beta^{(2)}$ | $\beta^{(3)}$ | $t^{(1)}$ | $t^{(2)}$ | $t^{(3)}$ | d |
|----|-------|-------|----------|------|-----------|-----------|---------------|---------------|---------------|---------------|-----------|-----------|-----------|------|
| Ag | 2.85 | 2.880 | 1.087 | 0.94 | 1.38 | 2.80 | 4.73 | 2.2 | 6.0 | 2.2 | 3.40 | 3.00 | 1.50 | 0.05 |
| Au | 2.85 | 2.880 | 1.803 | 1.00 | 1.53 | 2.80 | 5.77 | 2.2 | 6.0 | 2.2 | 2.90 | 1.64 | 2.00 | 0.05 |

To describe a binary alloy, an additional 13 parameters along with the constituent unary parameters are necessary. Among them, $E_c$, $r_e$, B, and d are related to the universal equation of state of the alloy. The atomic electron density scaling factor $\rho^0$ belongs to the electron density, and the remaining eight parameters, four $C_{min}$ and four $C_{max}$, comprising different combinations of Au and Ag as i, j, k atoms, are responsible for the many-body screening. Parameter d is kept same for both Au and Ag. Among the 13 parameters, 12 are parametrised in this study using DFT calculation and experimentally observed data as displayed in the table below-

Table 2: Parameters for the MEAM potential of Ag-Au alloy system.

| | Selected value | Procedure for determination |
|---|---|---|
| $E_c$ | -3.41 eV | DFT |
| $r_e$ | 2.89 A | Experimental |
| B | 147GPa | DFT |
| $C_{min}(Ag-Au-Ag)$ | 1.38 | Calibration against TEC |
| $C_{min}(Au-Ag-Au)$ | 1.53 | Calibration against TEC |
| $C_{min}(Ag-Ag-Au)$ | 1.45 | Calibration against TEC |
| $C_{min}(Ag-Au-Au)$ | 1.46 | Calibration against TEC |
| $C_{max}(Ag-Au-Ag)$ | 2.80 | Calibration against TEC |
| $C_{max}(Au-Ag-Au)$ | 2.80 | Calibration against TEC |
| $C_{max}(Ag-Ag-Au)$ | 2.80 | Calibration against TEC |
| $C_{max}(Ag-Au-Au)$ | 2.80 | Calibration against TEC |
| $\rho_0$ | 1 | Both being FCC |

Table 3 shows the *a*, $E_c$, B, TEC determined from MD using the parametrized MEAM potential, and compares the data with DFT and experimental results. The variation of the four properties for different compositions of Au and Ag, obtained using MEAM potential, is plotted with the reference data in Figure 2. The figure shows that the parametrized MEAM potential is applicable throughout the entire composition range of Au-Ag alloy.

Table 3: Comparison between MEAM, ab-initio, and experimental predictions (Neighbours and Alers, 1958; Okamoto and Massalski, 1983; Owen and Yates, 1933) for the cohesive energy, lattice parameter, TEC, and bulk modulus of pristine Au, Ag, and their alloy. a is the equilibrium lattice constant, B is the bulk modulus, $E_c$ is the cohesive energy, $\varepsilon$ is thermal expansion co-efficient

| Composition | Properties | DFT(PP) | | DFT/LAPW | MEAM | Experimental |
|---|---|---|---|---|---|---|
| | | USPP | PAW | | | |
| Ag | a(A) | 4.1586 | 4.1583 | 4.0491 | 4.0870 | 4.0862 |
| | $E_c$ | -2.550 | -2.548 | -2.826 | -2.849 | -2.95 |
| | **B** | 79.2 | 85.9 | 103 | 103.13 | 109 |
| | $\varepsilon$ | - | - | - | 17.2 | 17.9 |
| Au | a(A) | 4.1682 | 4.1676 | 4.1188 | 4.0787 | 4.0784 |
| | $E_c$ | -3.035 | -3.029 | -3.877 | -3.929 | -3.81 |
| | **B** | 139.4 | 137.6 | 197 | 165.73 | 180 |
| | $\varepsilon$ | - | - | - | 10.72 | 13.1 |
| Ag-Au | a(A) | 4.1551 | 4.1548 | 4.0780 | 4.0765 | 4.077 |
| | $E_c$ | -2.858 | -2.854 | -3.419 | -3.470 | - |
| | **B** | 115.3 | 111.3 | 147 | 133.09 | - |
| | $\varepsilon$ | - | - | - | 12.57 | 13.8 |

The table 3 and figure 2 show that the lattice parameters obtained from parametrized MEAM potential better represent the experimental data than DFT. The bulk modulus and cohesive energy of both DFT and MD are in agreement with experiments.

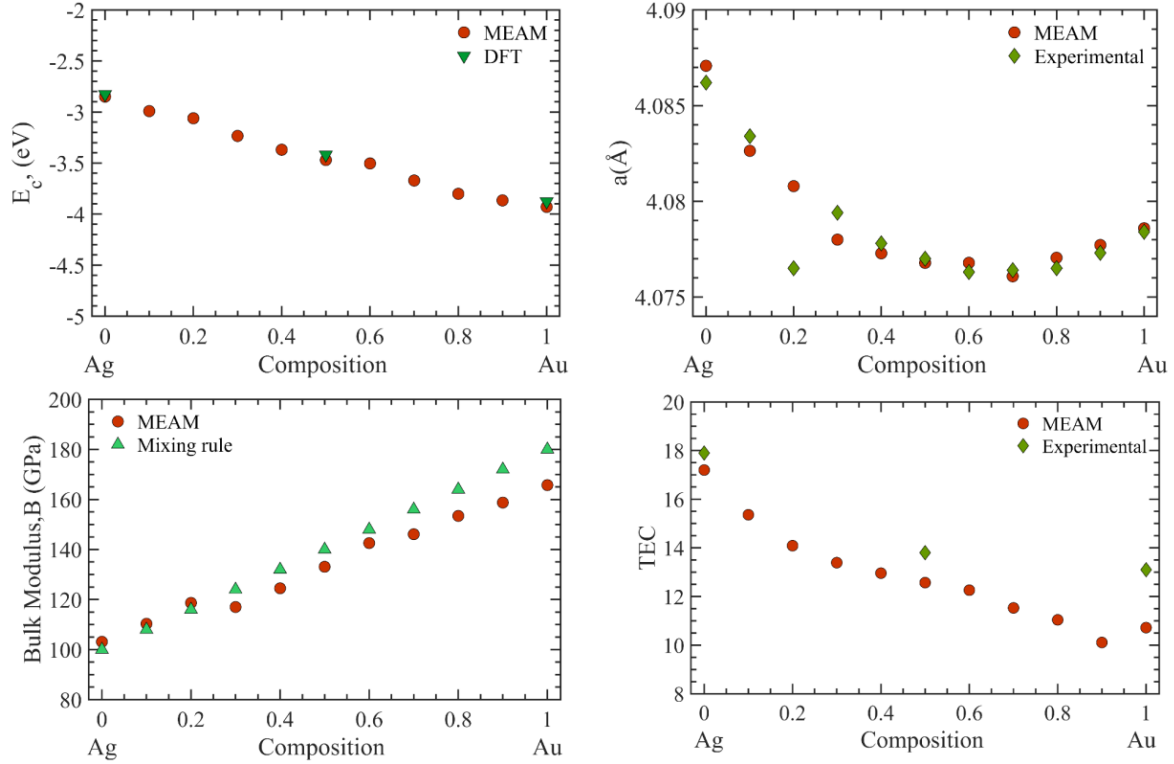

Figure 2: Comparison of (a) Cohesive energy (MEAM and DFT), (b) lattice parameter (MEAM and experimental), (c) bulk modulus (MEAM and mixing rule), and (d) TEC (MEAM and Experimental(Okamoto and Massalski, 1983; Owen and Yates, 1933)), for different compositions of Au in Ag.

### 3.2. Effects of temperature and NW composition

We investigate how the material properties of the Au, Ag, Au-Ag, and Ag-Au NWs modulate with temperature under tension. Figure 3 represents the tensile behavior of the NWs at different temperatures. Pristine Au and Ag NWs undergo elastic deformation up to 7% strain and exhibit an ultimate tensile strength (UTS) of 4.53 GPa and 4.75 GPa and Young's Modulus (YM) of 105.06 GPa and 94.23 GPa, respectively, at 300K. The UTS and YM from the MD simulations agree with previous studies(Deb Nath, 2014; Joshi et al., 2019; Matthew T. McDowell et al., 2008; Matthew T McDowell et al., 2008; Park and Zimmerman, 2005). With increasing temperature from 300K-600K, UTS of Au and Ag NWs reduces by 9.5% and 15.3%, respectively. Thermal weakening is more evident in Ag than Au. Au-Ag and Ag-Au CSNWs exhibited a UTS of 4.71 GPa and 4.65 GPa and a YM of 104.50 GPa and 90.36 GPa, respectively, at 300K. From Figure 1, we see that the core-shell ratio is 1:1.47. Hence, the mechanical properties of the CSNWs are closer to the constituent material of the shell. With increasing temperature, more significant reduction is observed in the UTS of Au-Ag CSNW (12.5%) than Ag-Au CSNW (11.5%). Both NWs and CSNWs experience early onset of plasticity at a lower strain with increasing temperature.

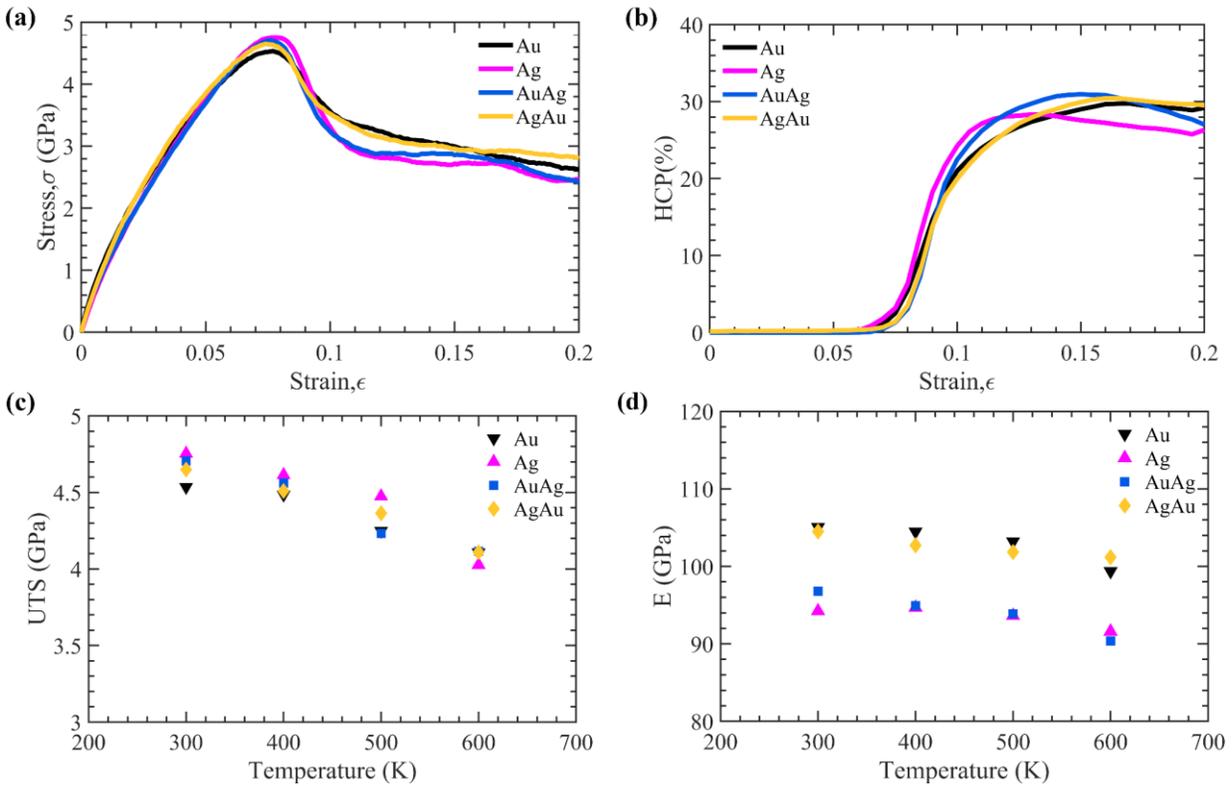

Figure 3: Mechanical properties of Au, Ag NWs and Au-Ag, Ag-Au CSNWs under tension where (a) shows the stress-strain response at 300K. (b) Shows variation of the percentage of atoms forming HCP lattice with strain at 300K. The variation of (c) UTS and (d) Young's modulus with increasing temperature from 300K to 600K is shown.

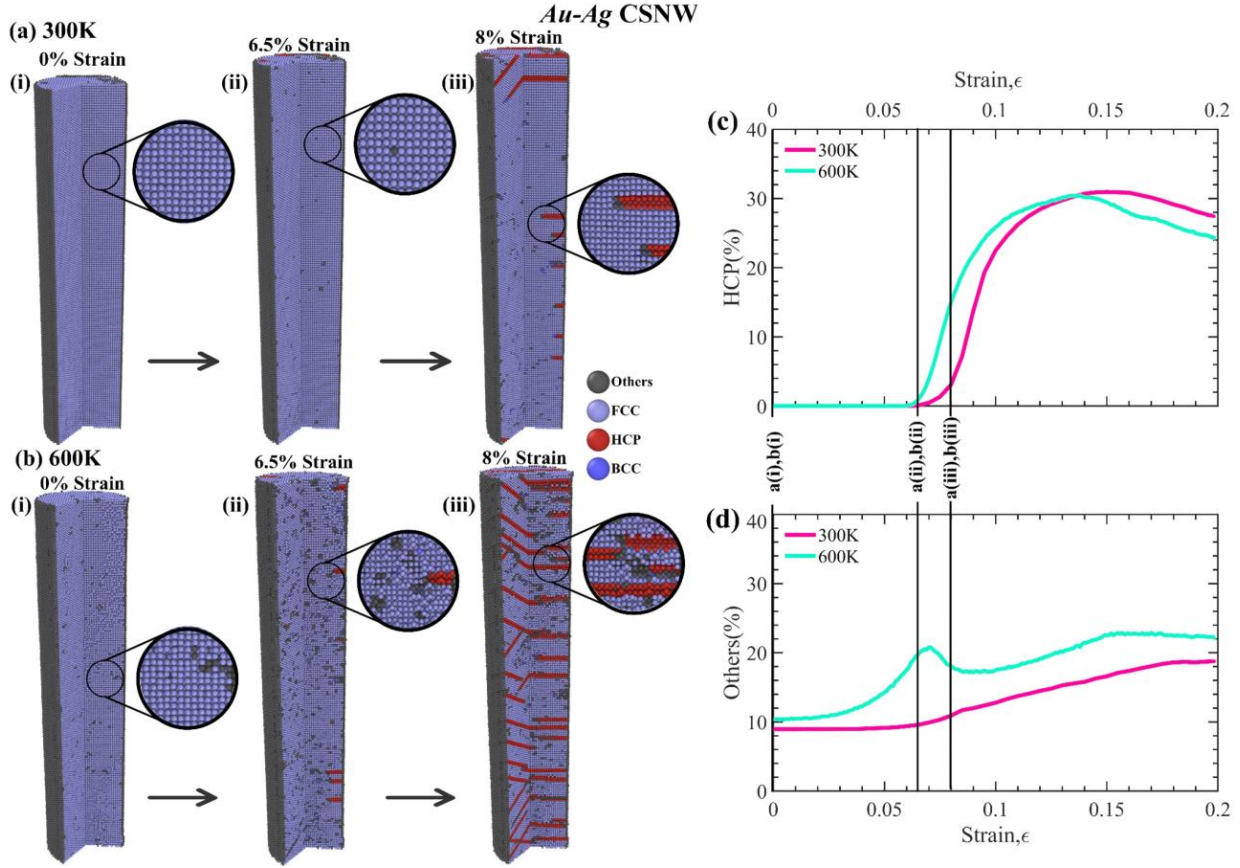

Figure 4: The effects of temperature on Au-Ag CSNW. CNA of Au-Ag CSNW at 0%, 6.5% and 8% strain at (a) 300K and (b) 600K. And the percentage of atoms which forms (c) HCP lattice and (d) non-crystalline (Others) lattice at 300K and 600K with increasing strain.

In Figures 4 (a, b) and 5 (a, b) the thermal softening phenomenon of Au-Ag and Ag-Au CSNWs is observed. Figures 4 (c, d) and 5 (c, d) show the percentage of non-crystalline structures and change of HCP plane formation with strain at 300K and 600K temperatures. The graphs show that increased temperature causes more non-crystalline structures to form. Both CSNWs show a spike in non-crystalline structure formation near their elastic limit. These phenomena occur due to increased kinetic energy in the lattice at higher temperature making the structure weaker.

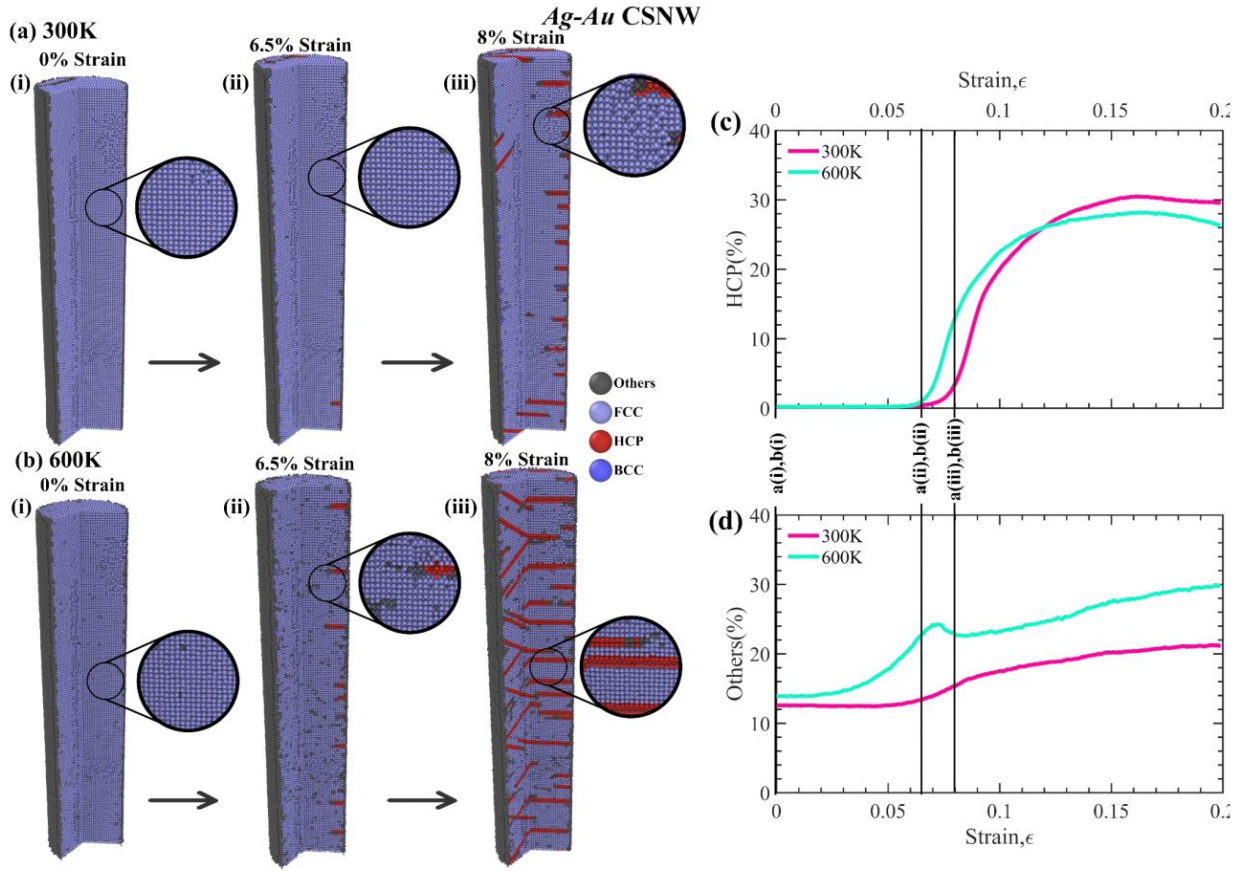

Figure 5: The effects of temperature on Ag-Au CSNW. CNA of Ag-Au CSNW at 0%, 6.5% and 8% strain at (a) 300K and (b) 600K. And the percentage of atoms which forms (c) HCP lattice and (d) non-crystalline (Others) lattice at 300K and 600K with increasing strain.

A rapid increase in HCP plane formation in Figures 4 (c, d) and 5 (c, d) indicates the onset of plastic deformation at a lower strain for a higher temperature. Primarily, the HCP formation at an increased temperature is higher. But upon crossing 13.5% strain and 12% strain for Au-Ag and Ag-Au CSNW, respectively, the HCP formation at higher temperature drops below that of lower temperature. This occurs due to a higher non-crystalline phase formation at higher temperatures, impeding HCP plane formation after a certain strain. These results are in line with previous studies on the effects of temperature on nanostructures(Faiyad et al., 2021; Munshi et al., 2019).

### 3.3. Effects of fatigue loading and role of dislocations on material strength

Figure 6 illustrates the response to cyclic loading of Au, Ag NWs and Ag-Au, Au-Ag CSNWs for ten cycles at 300K. The NWs and CSNWs were subjected to -15% to 15% strain from their original length. In the first cycle, all four structures deform elastically up to about 7-9% strain. Then plastic deformation continues up to 15% strain. Then, the loading direction is reversed. At about 8-10% compressive strain, all four structures show a peak compressive stress and start to flow up to 15% compression. This behaviour is inline with previous MD studies of metal alloy NWs under compressive loading (Kardani and Montazeri, 2020; Tang, 2012).

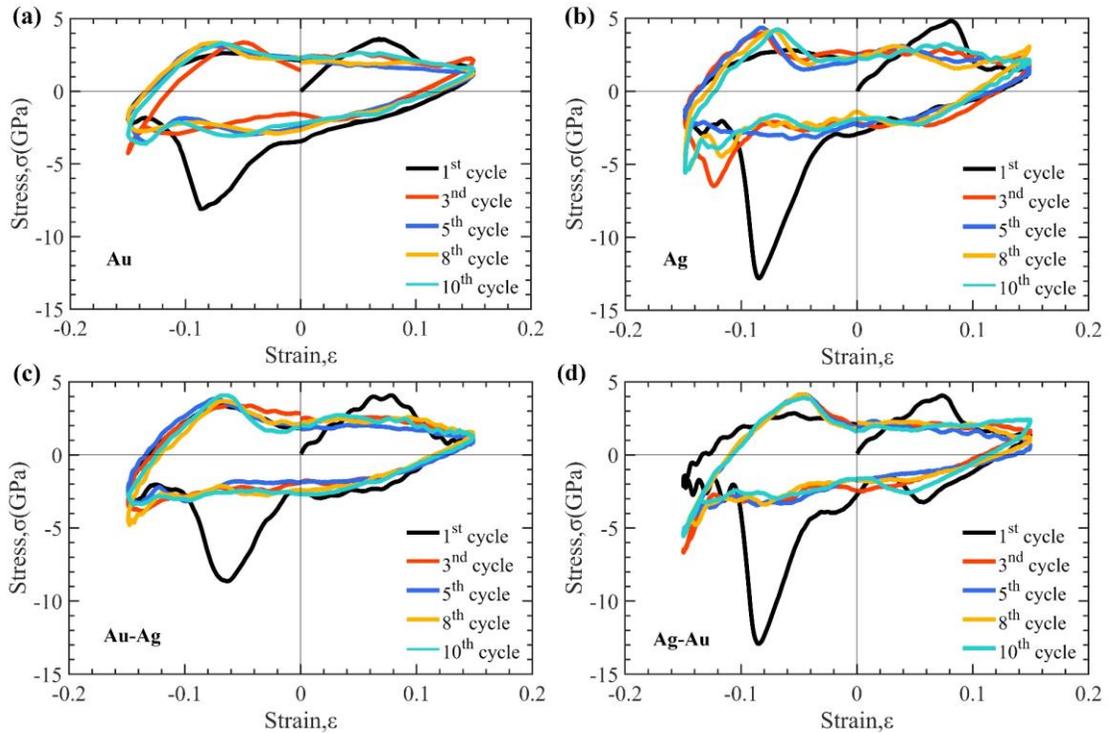

Figure 6: Variation of Stress with cyclic increase and decrease of strain in the range of 15% to -15% for (a) Au, (b) Ag NWs and, (c) Au-Ag, (d) Ag-Au CSNWs upto 10 cycles.

Figure 6 shows that the UTS and ultimate compressive stress (UCS) of pristine Ag NW (4.12 GPa and -12.93 GPa, respectively) is the highest and Au NW (3.07 GPa and -8.27 GPa, respectively) shows the lowest. The Au-Ag (3.92 GPa and -8.75 GPa, respectively) and Ag-Au (4.02 GPa and -12.85 GPa, respectively) CSNWs show UTS and UCS values between those of Au and Ag NWs. The initial defect free structures, typical for MD, partly explain why the structures behave differently in the first cycle from the subsequent cycles. During initial compressive loading, a significant compressive peak is observed around -8% to -10% strain which can be attributed to sudden drop of dislocation density as seen in Figure 7 (a, c) and also due to the <1 1 0> orientation of the NWs and CSNWs which has been previously observed in similarly oriented FCC alloy NWs (Mojumder, 2018).

From the 2$^{nd}$ cycle, due to stress retention and plastic deformations, the NWs and CSNWs contain twining and dislocation at 0% strain. The structures now reach a dislocation and twining free state around -13% strain, indicating that the structures are acting like shorter NWs. The UTS of the subsequent cycles lying in the -4% to -8% range as seen in Figure 7 reinforces this finding. Here, all four structures have almost no dislocations and twins at -15% strain in the 5$^{th}$ and 10$^{th}$ cycle, whereas for the first cycle, dislocation-free structures are observed at 0% strain. The percentage change of UTS can quantify the strength retention of the structures during the fatigue test. In Au, Ag, Au-Ag, and Ag-Au, a decrease of 2.4%, 8.9%, 0.4%, and 7.2%, respectively, in the UTS is observed from the 1$^{st}$ to 10$^{th}$ fatigue cycle.

Plastic deformation in metallic NWs typically occurs through slips or twinning. Slips involve the movement of full dislocations and surface step formation, whereas twinning occurs through the movement of partial dislocations in FCC NWs. The deformation mechanism of <110> oriented FCC NWs, used for all four structures in this paper under tensile and compressive loading, is presented by Harold et al.(Park et al., 2006) From Figure 8 and Figure 9 it is observed that Au and Ag NWs show deformations primarily through Shockley partial dislocation in tension with twinning-like structure. While in compression, both NWs deform through full and partial dislocations, creating surface steps and partial dislocations with no twinning. Both of the responses are in good agreement with Harold et al(Park et al., 2006). Au and Ag NWs showing twinning-like responses with partial dislocation in tension and full dislocation slip in compression can be explained through the respective Schimd factors of the two processes. The Schimd factor for slip and tensile-twin is close to 0.5 in the case of <110> NWs. Therefore, the twinning is to be favored in tension and slip in compression. However, twinning is not observed in both cases. Instead, parallel stacking faults are observed. This phenomenon can be attributed to the lattice orientation of <110> NWs and the associated crystallographic constraints. The tensile loaded <110> NWs cannot form low energy <111> side surfaces due to twinning because of crystallographic constraints. The formation of twins requiring the nucleation of sequential partial dislocations on adjacent planes is not energetically favorable compared to dispersed stacking faults. Thus, the latter is observed in the tensile loaded <110> NWs. These twinning-like formations are more reversible than full dislocation slip because dislocation slips lead to significant distortion and shearing of the NWs. This results in the NWs deforming more reversibly during tension than in compression.

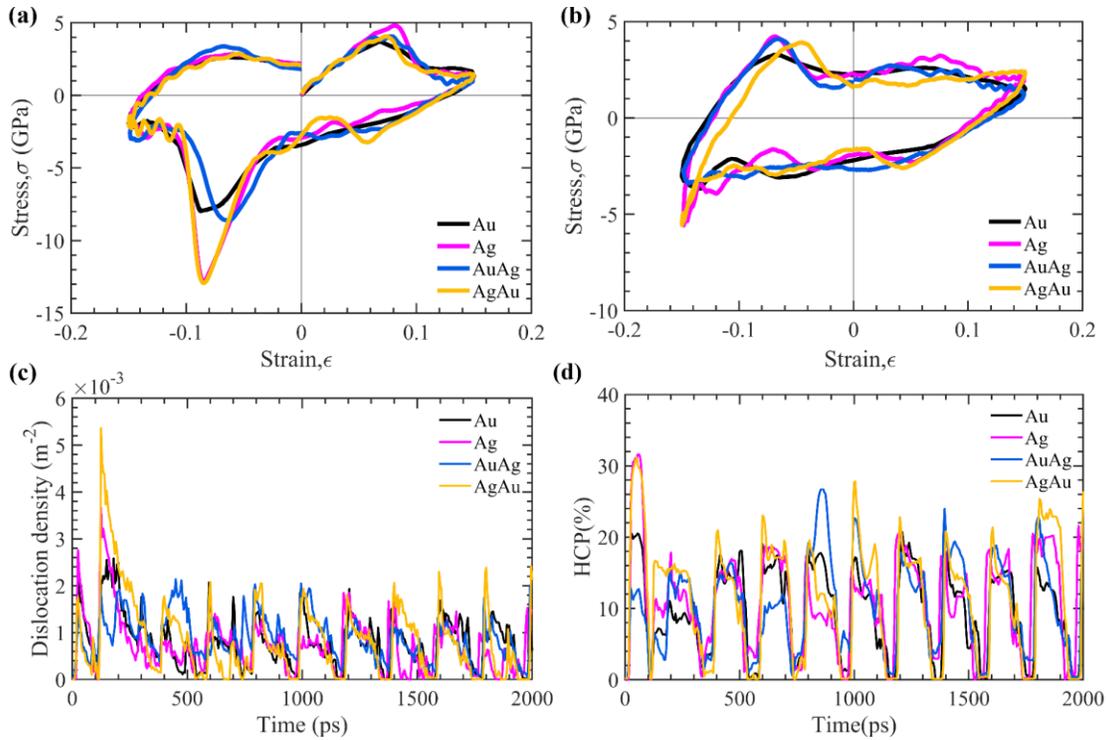

Figure 7: The variation of stress with cyclic increase and decrease in strain in the range of 15% to -15% for Au, Ag NWs and Au-Ag, Ag-Au CSNWs for (a) 1$^{st}$ and (b) 10$^{th}$ cycles. (c) Represents the dislocation densities of the structures and (d) shows the percentage of atoms which forms HCP lattice with cyclic strain for 10 cycles where each cycle lasts for 200ps.

Figure 7 (c, d) shows that Ag exhibits more dislocations and reversibility than Au Nw. Au doesn't exhibit as much dislocation and twin plane formation showing poor stress reversibility. Au NW shows a decline in its ability to generate HCP faults, dislocations, and reversal of defects in subsequent cycles. Degradation in Ag NWs reversibility is not as prominent as Au NW.

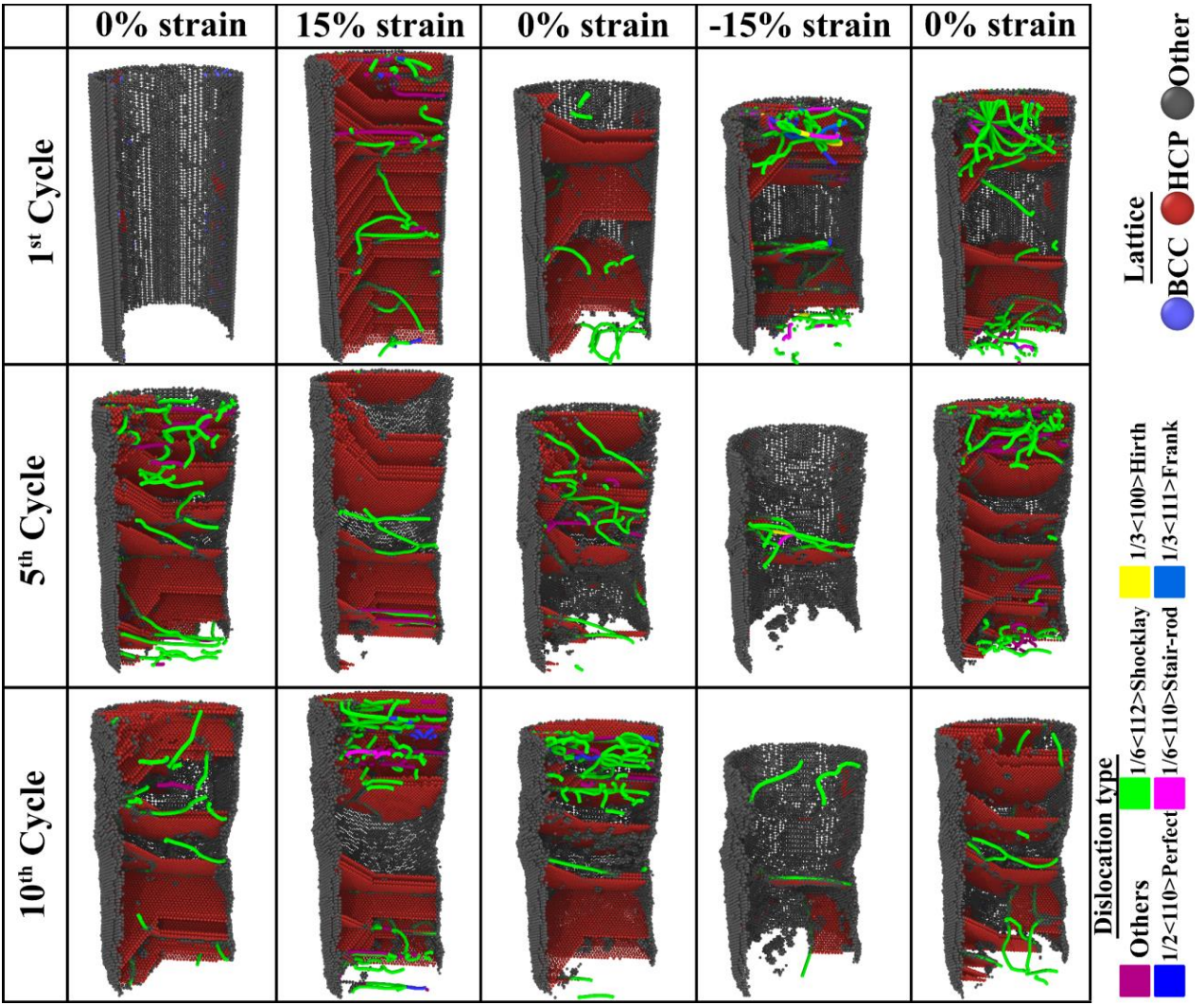

Figure 8: A section of Au NW where FCC atoms have been removed to observe the stacking fault, twins and dislocations at 1st, 5th and 10th cycles.

Figure 10 and 11 shows that the CSNWs have a similar mechanism with twinning-like stacking fault accompanied by partial dislocations during tension and with full and partial dislocations during compression, as with Au and Ag NWs. But variability is observed in terms of dislocation nucleation and propagation between the NWs and CSNWs. This variability in dislocation behavior and stacking fault formation is displayed by plotting the dislocation density and HCP percentage with time in Figures 7 (c, d).

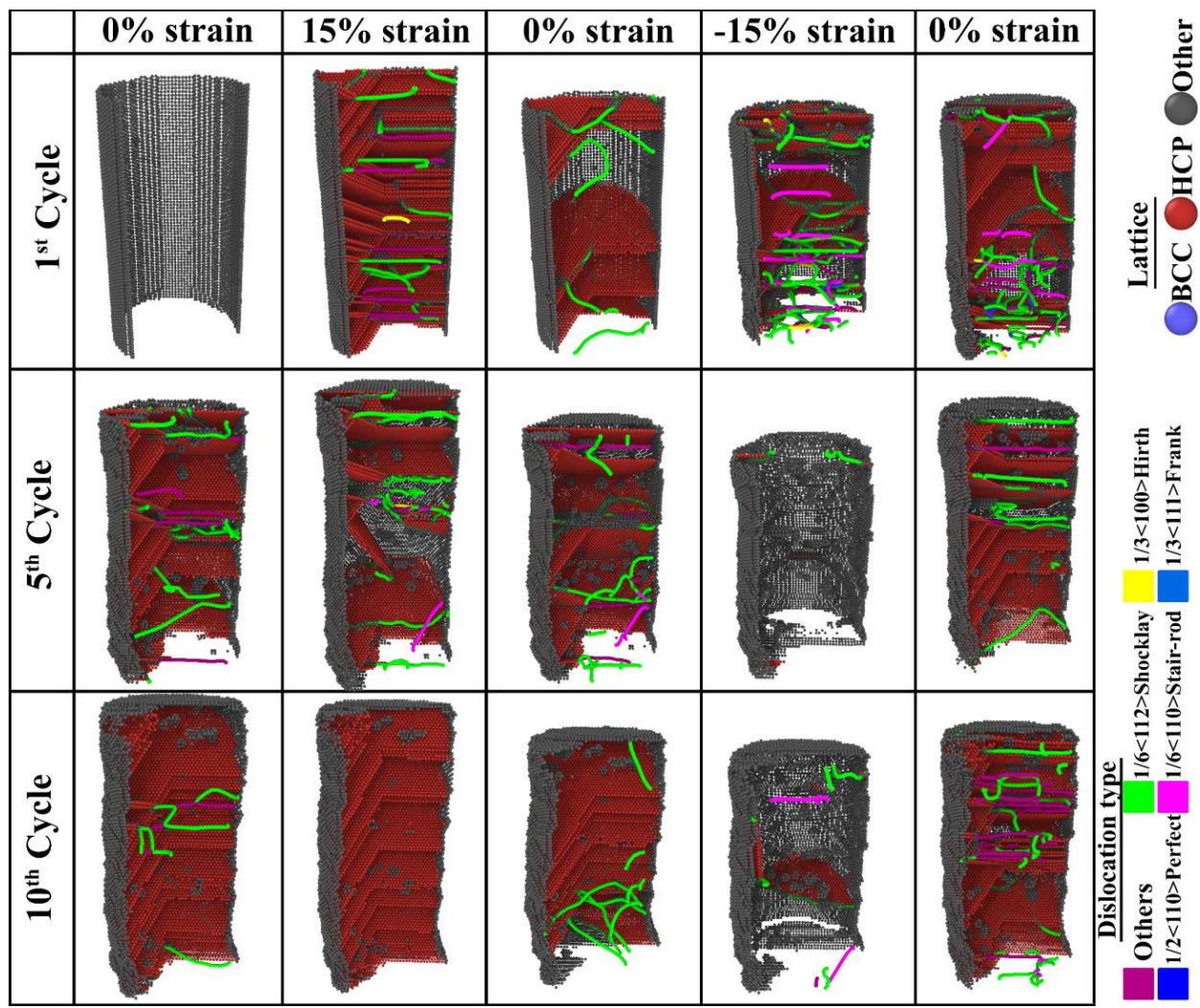

Figure 9: A section of Ag NW where FCC atoms have been removed to observe the stacking fault, twins and dislocations at 1st, 5th and 10th cycles.

The Ag-Au CSNW has the steadiest fluctuation in HCP formation and dislocation density. This structure consistently gives rise to the most dislocations and HCP planes during tension accompanied by a near-complete reversal of its structure under compression by complete annihilation of dislocation and stacking faults. The complete reversal from the strained structure of Ag-Au can be seen from Figure 11 at -15% strain (5th and 10th cycle) which shows no dislocation and HCP planes. Moreover, the Ag-Au structure shows high dislocation density and HCP planes during tensile loading. This behavior is a testament to the excellent fatigue resistance of Ag-Au

CSNW. The Au-Ag CSNW also shows steady peaks of dislocation densities and HCP planes throughout the 10 cycles, as represented in Figure 10. However, in comparison to Ag-Au, the properties of Au-Ag deteriorate by a larger margin.

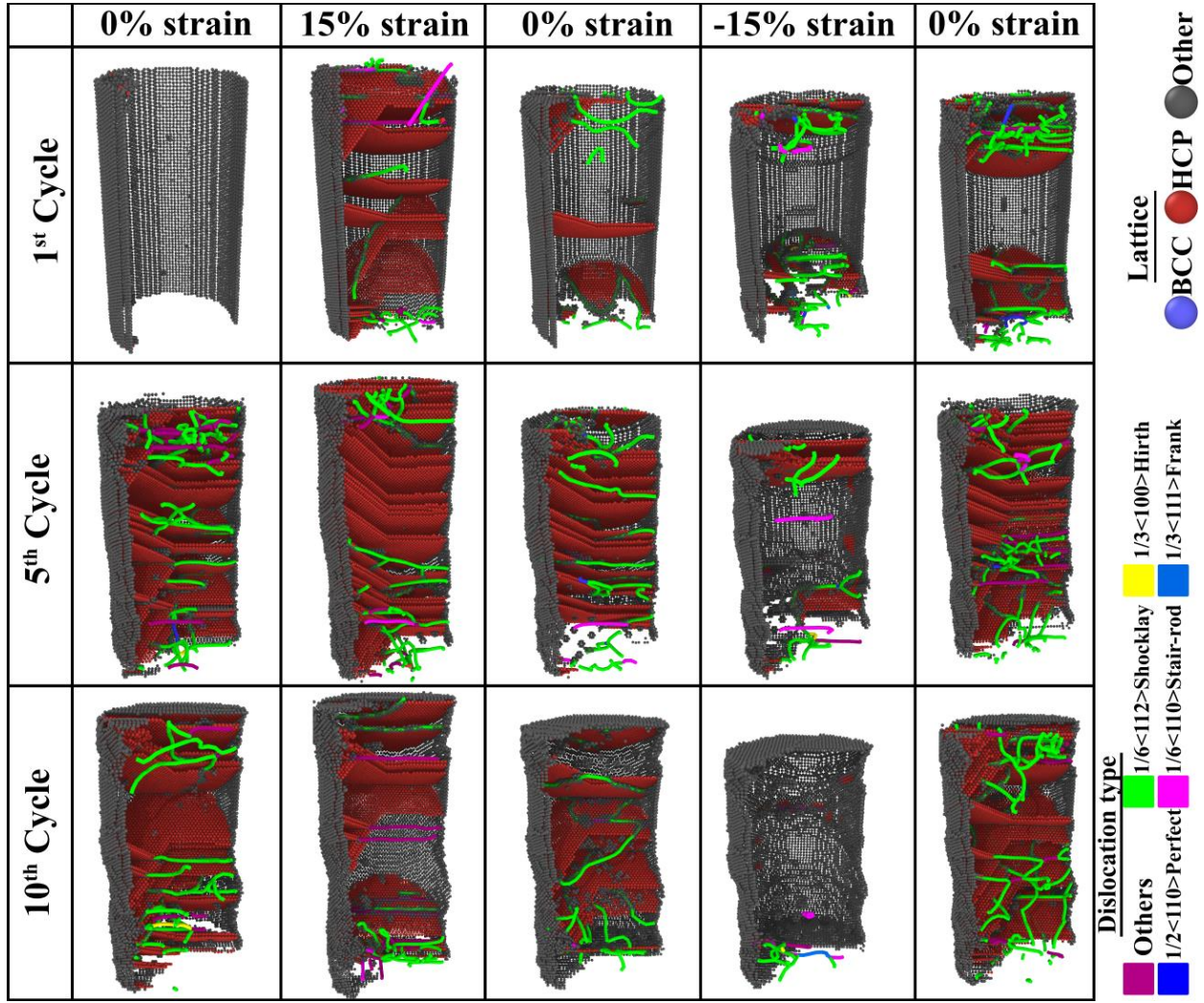

Figure 10: A section of Au-Ag CSNW where FCC atoms have been removed to observe the stacking fault, twins and dislocations at 1st, 5th and 10th cycles.

The Au-Ag CSNW retains some dislocations and HCP plane on full reversal of strain. This phenomenon is depicted in Figures 7 (c, d) where non-zero dislocation density and HCP percentage are observed in between two cycles. This notion is further reinforced by Figure 10 where dislocations and HCP planes are observed in the 5th and 10th cycle at -15% strain.

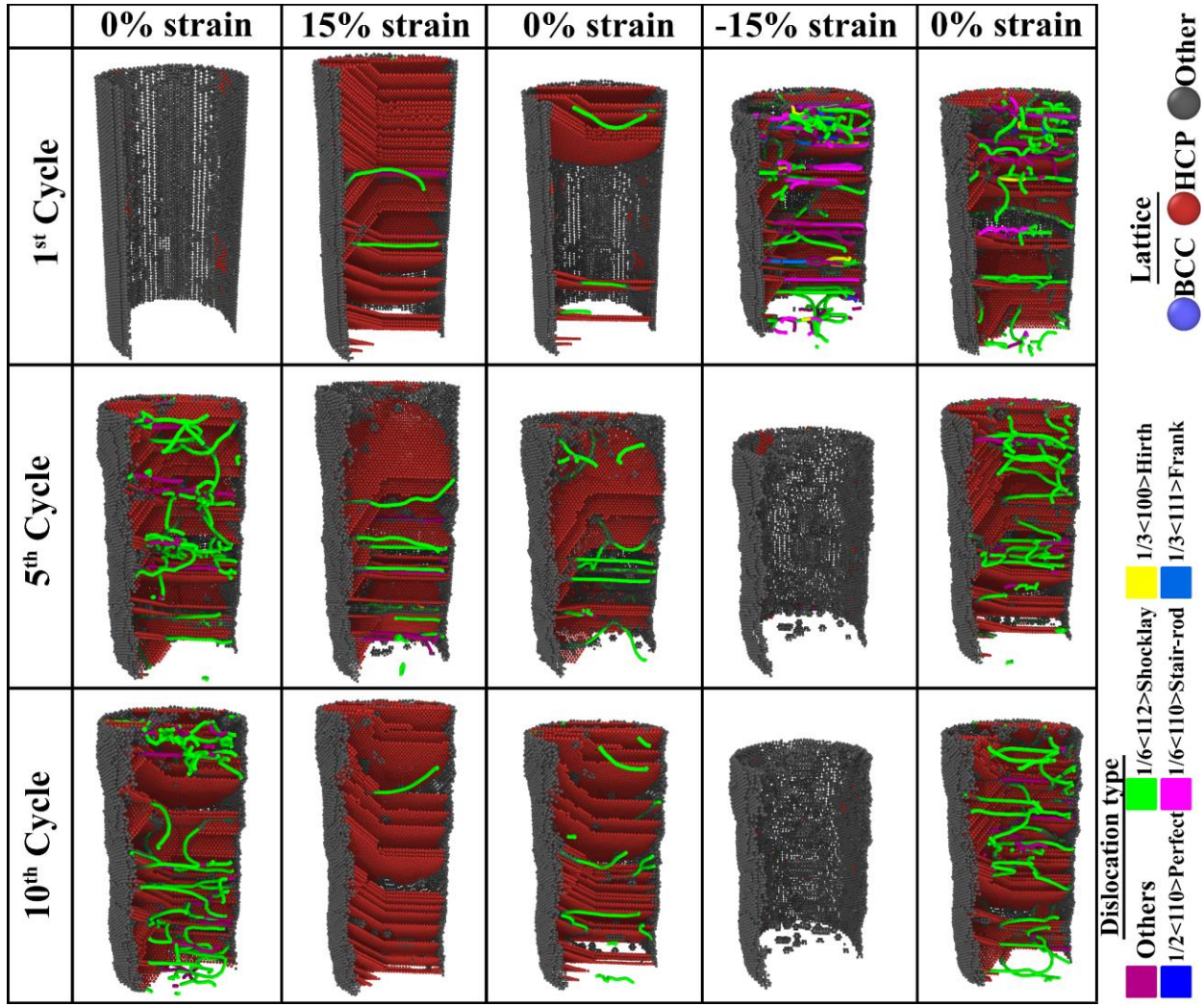

Figure 11: A section of Ag-Au CSNW where FCC atoms have been removed to observe the stacking fault, twins and dislocations at 1st, 5th and 10th cycles.

Unreversed HCP formations, slip planes and dislocations have been shown to influence the initiation and propagation of micro-cracks which in turn shortens the service life during fatigue loading(Fan et al., 2016). From our study it has been observed that, Ag-Au can inhibit this micro-crack initiation and propagation by achieving complete reversal of dislocation, HCP plane and stress during repeated unloading which can lengthen their operational life. Au-Ag also displays some level of dislocation and stress mitigation. This annilation process is absent in their pristine counterparts. Hence, both the CSNWs are a better candidate for stretchable electronics, bio-electronics and NEMS application where good fatigue resistance is paramount.

## 4. Conclusions

In this study, we develop a MEAM interatomic potential for Ag-Au alloy to investigate the mechanical behavior of Au and Ag CSNW systems. Applications of Au and Ag CSNWs under tensile and fatigue loading, which can affect their applicability in stretchable and biocompatible electronics, is elucidated in this study. We find that, the AEFP DFT scheme yielded more accurate-physical properties of Au, Ag and Au-Ag than those obtained from pseudo-potential schemes. The MEAM potential parametrized exhibits good agreement with DFT and experimental data for all compositions of Au-Ag alloys. In MD, pure Ag NW displayed the higher UTS than pure Au NW under tensile load. The UTS of CSNWs are between that of Ag and Au NW. For the selected core-shell ratio, the UTS of the CSNWs approachs that of the constituent shell material because of its higher volume fraction. Therefore, Au-Ag CSNW shows higher UTS than Ag-Au CSNW.

With increasing temperature, mechanical properties of all four structures deteriorate due to thermal softening. This effect is more prominent in Ag NW than Au NW. Due to the higher volume fraction of shell, the CSNWs followed the deterioration trend of the material in the shell. Thus, Ag-Au CSNW supersedes Au-Ag CSNW in tensile strength at higher temperature.

During cyclic loading, among the four structures, Ag-Au CSNW showed the best reversibility in terms of dislocation and stacking faults, twin formation and annihilation. It exhibits higher dislocation and stacking fault formation during loading and complete reversal of them during unloading among the four structures for several cycles. Whereas, with the other three structures develop some form of softening, resulting in a gradual decrease of dislocation and stacking fault formation and their retrieval during cyclic loading.

## 5. Acknowledgements

The authors of this paper would like to acknowledge Multiscale Mechanical Modeling and Research Network (MMMRN) and the Department of Mechanical Engineering, Bangladesh University of Engineering and Technology (BUET) for the technical support to conduct the research. MMI acknowledges start up funds from Wayne State University.